\documentclass[aps,pre,twocolumn,superscriptaddress,showkeys,showpacs,amsmath,amssymb, longbibliography,nofootinbib ]{revtex4-1}

\usepackage{graphicx}
\usepackage{dcolumn}
\usepackage{bm}
\usepackage{hyperref}

\newcommand{\ket}[2][]{{|#2\rangle_{#1}}}
\newcommand{\bra}[2][]{{}_{#1}\langle #2|}
\def\<{\langle}
\def\>{\rangle}

\def\ben{\begin{eqnarray}}
\def\een{\end{eqnarray}}

\begin{document}

\title{Out-of-time-ordered correlation functions in open systems: A Feynman-Vernon influence functional approach}

\author{Jan Tuziemski}\email{jan.tuziemski@fysik.su.se}
 \altaffiliation[On leave from ]{Department of Applied Physics and Mathematics, Gda\'nsk University of Technology}
 \affiliation{Department of Physics, Stockholm University, AlbaNova University Center, Stockholm SE-106 91 Sweden}
 \affiliation{Nordita, Royal Institute of Technology and Stockholm University, Roslagstullsbacken 23, SE-106 91 Stockholm, Sweden}

\begin{abstract}
	Recent theoretical and experimental studies have shown significance of quantum information scrambling (i.e. a spread of quantum information over a system degrees of freedom) for problems encountered in high-energy	physics, quantum information, and condensed matter. Due to complexity of quantum many-body systems it is plausible that new developments in this field will be achieved by experimental explorations. Since noise effects are inevitably present in experimental implementations, a better theoretical understanding of quantum information scrambling in systems affected by noise is needed. To address this problem we study indicators of quantum scrambling -- out-of-time-ordered correlation functions (OTOCs) in open quantum systems. As most experimental protocols for measuring OTOCs are based on backward time evolution we consider two possible scenarios of joint system-environment dynamics reversal: In the first one the evolution of the environment is reversed, whereas in the second it is not. We derive general formulas for OTOCs in those cases as well as study in detail the model of a spin chain coupled to the environment of harmonic oscillators. In the latter case we derive expressions for open systems OTOCs in terms of Feynman-Vernon influence functional. Subsequently, assuming that dephasing dominates over dissipation, we provide bounds on open system OTOCs and illustrate them for a spectral density known from the spin-boson problem. In addition to being significant for quantum information scrambling, our results also advance understating of decoherence in processes involving backward time evolution. 
\end{abstract}

\date{\today}
\maketitle


\section{\label{sec:intro}Introduction}

Dynamics of a quantum many-body system leads to a spread of quantum information across its degrees of freedom. As a consequence localized states become inaccessible to local measurements. This phenomenon, refereed to as scrambling, has recently become a vivid area of research joining different fields of physics such as quantum information theory\cite{Entropicuncertaintyrelations,OTOCEntanglement}, quantum field theory \cite{MSS2016-bound,RS2016-Butterfly_effect_Lieb_Robinsnon}, and condensed matter\cite{OTOCO(N),OTOCLuttiger}. Studies of quantum information scrambling allowed to gain new insights into problems such as thermalisation (see e.g. \cite{Sch2017-slow_scrambling_disordered}) or many-body chaos \cite{LyapunovOTOC} in quantum systems. Scrambling can be diagnosed by unusual correlation functions called out-of-time-ordered correlators (OTOCs), which for two operators $V$ and $W$ read
\begin{equation}
    F_t(V,W) = \langle W_t^{\dagger} V^{\dagger} W_t V \rangle_{\rho} = Tr\left(W_t^{\dagger} V^{\dagger} W_t V \rho \right),
\end{equation}
where $W_t = e^{i t H} W e^{-i t H}$ and $H$ is a Hamiltonian of a considered system. Contrary to standard correlators, a measurement of OTOCs involves backward time evolution that must be applied twice to the investigated system. Backward time evolution makes OTOCs similar to Loschmidt echo (LE) \cite{Loschmidtecho}, however in the latter only one imperfect reversal of dynamics is applied, and no measurements in between are made. The aim of OTOCs is to measure how quickly two, initially commuting, operators $W$ and $V$ cease to commute (OTOCs can be seen as a state-dependent version of Lieb-Robinson bounds \cite{RS2016-Butterfly_effect_Lieb_Robinsnon}), whereas LE aims to capture sensitivity of a system’s evolution to perturbations. For a more detailed discussion regarding relations between OTOCs and LE see \cite{Zurek}.  The crucial feature of OTOCs is their time dependence: the faster an OTOC decays, the shorter is the scrambling time, which indicates onset of quantum chaos in the considered system. 

So far quantum information scrambling has been investigated mostly in the isolated system setting. OTOCs were used to characterize chaotic behavior of several types of systems \cite{OTOCO(N),Sch2017-slow_scrambling_disordered,LSSNBR2018_scrambling_Dicke_model,ScramblingManyBody, OTOCLuttiger, OTOCManyBodyLocal,OTOCChargeConservation}  and a bound on their decay rate was conjectured \cite{MSS2016-bound}. Several experimental scenarios to measure OTOCs were proposed \cite{SBSchCH2016-Scrambling_measurement, Yaoetal-Scrambling-inteferometric-measurement}  and results of first experiments were reported 
\cite{Garttner2017-scrambling-epxperiment1, Lietal2018-scrambling_experiment2, WRC2018-scrambling_experiment3, MAAG2017-scrambling_experiment4,scramblingteleportation}.
Moreover, links between OTOCs and thermodynamics  \cite{YH2017-Jarzynski, CG2017-Thermodynamics_OTOC, TST2018-OTOC_Fluctuation_Disipattion_theorem}, quasi-probabilities \cite{YHSD2018-Quasiprobability}, and quantum information \cite{OTOCEntanglement, Entropicuncertaintyrelations} were investigated. 

Although very convenient, the notion of an isolated system constitutes an idealization. In real situations, such as experimental apparatus, all systems are open -- due to interaction with the environment they are influenced by external noise. Physics of open quantum systems is qualitatively different than that of closed ones: Open systems decohere losing their quantum coherence as well as energy to the environment. Therefore one expects that an interaction with the environment will affect the spread of quantum information in the considered system. Yet the issue how the coupling to the environment influences quantum information scrambling remains vastly unexplored.

Only few works address scrambling in open quantum systems: a master equation was derived \cite{SGG2018-Master}, a measurement protocol was proposed \cite{YY2018-DisentanglingScrambling} as well as numerical studies were performed \cite{YHS2018-Resiliance,ZHCh2018-ScramblinDissipation}. Thus the aim of this work is to advance understanding of open system OTOCs. The main used tool is the Feynman-Vernon influence functional \cite{FV1963}, which allows to capture an influence of the environment on the studied system. This approach proved useful both from practical \cite{CL1983-BrownianMotionFV,LCDF+1987-DynamicsDissipativeTwoLevel,Weiss-Book} as well as fundamental point of view \cite{GMH1993-ClassicalEquationsQuantumSystems,DH1992-DecoherenceFunctional,GJKK+1996-book}, and still provides insights into problems encountered in fields such as open systems \cite{MSHP2017-SpinBosonError}, quantum thermodynamics \cite{Aurell2017-OnWorkAndHeat, AE2017-vonNeumanEntropyFV, Aurell2018-CharacteristicQuantumHeat, FQ2018-PathIntegralsQuantumThermo,  CSBSW2015-FunctionalHeatExchangeFV0, CSSW2016-EnergyExchange, AurellMontana} or quantum computing \cite{Aurell2018-EstimationofError}. As we show here it encapsulates decoherence of OTOCs in therms of the microscopic parameters of the considered model, allows to gain better insight into differences between the two considered backward time evolution schemes as well as has useful applications e.g. can be used to  bound the difference between OTOCs in isolated systems and their open-system counterparts. The results of this paper not only contribute  to the particular problem of open system OTOCs, but also advance our understanding decoherence in processes involving backward time evolution.

The paper is organized as follows. In Section \ref{sec:OSOTOC} possible schemes of backward time evolution are discussed and corresponding expressions for open system OTOCs are derived. Subsequently, in Section \ref{sec:OTOC_spinchain}, we provide expressions for open system OTOCs in terms of Feynman-Vernon influence functional for a class of spin chain systems interacting with a bosonic environment. In Section \ref{sec:applications} the obtained results are analyzed and some applications of the proposed approach are discussed. In particular, a lower bound on open system OTOCs is provided. 
 Summary and open questions are provided in Section \ref{sec:discussion}. Details of derivations are presented in Appendix \ref{sec:appecolution}.   

\section{OTOCs in open systems \label{sec:OSOTOC}}
In the standard treatment of open systems it is assumed that environmental degrees of freedom are out of control, so that they should be traced out from the description of the problem. This, together with the fact that the environment couples to the system, leads to the well-known phenomenon of decoherence \cite{GJKK+1996-book}. Application of the open system paradigm to the quantities involving backward time evolution, such as OTOCs, leads to two possibilities regarding behavior of the environment. In the first case one assumes that the joint dynamics of the system and the environment can be reversed perfectly. Then OTOCs will be affected by decoherence, which is caused by limitations of a measurement apparatus that is not capable of measuring all environmental degrees of freedom, which interacted with the system, and backward time evolution is simply $U_{SE}^{\dagger}=e^{iH_{SE}t}$, with a joint Hamiltonian $H_{SE}= H_S + H_E + H_{S:E}$ describing an evolution of the system - $H_S$, the environment - $H_E$ and an interaction between them -  $H_{S:E}$. As negation of the total Hamiltonian effectively implements reversal of the evolution, we will refer to this case as to the full backward time evolution (FBTE) case. However, due to complexity of the process, it may be impossible to reverse dynamics of the environment. In such a case only backward time evolution of the system can be implemented, which reads $U_{S^{\dagger}E} = e^{i(H_S-H_E-H_{S:E})t}$ (for a detailed discussion of possible backward time evolution schemes for open systems see \cite{AurellZZ} ).  We will refer to this case as to the partial backward time evolution (PBTE) case (This corresponds to the canonical time reversal in stochastic thermodynamics \cite{Gawedzki2008}).

\begin{figure}

\includegraphics[scale=0.22]{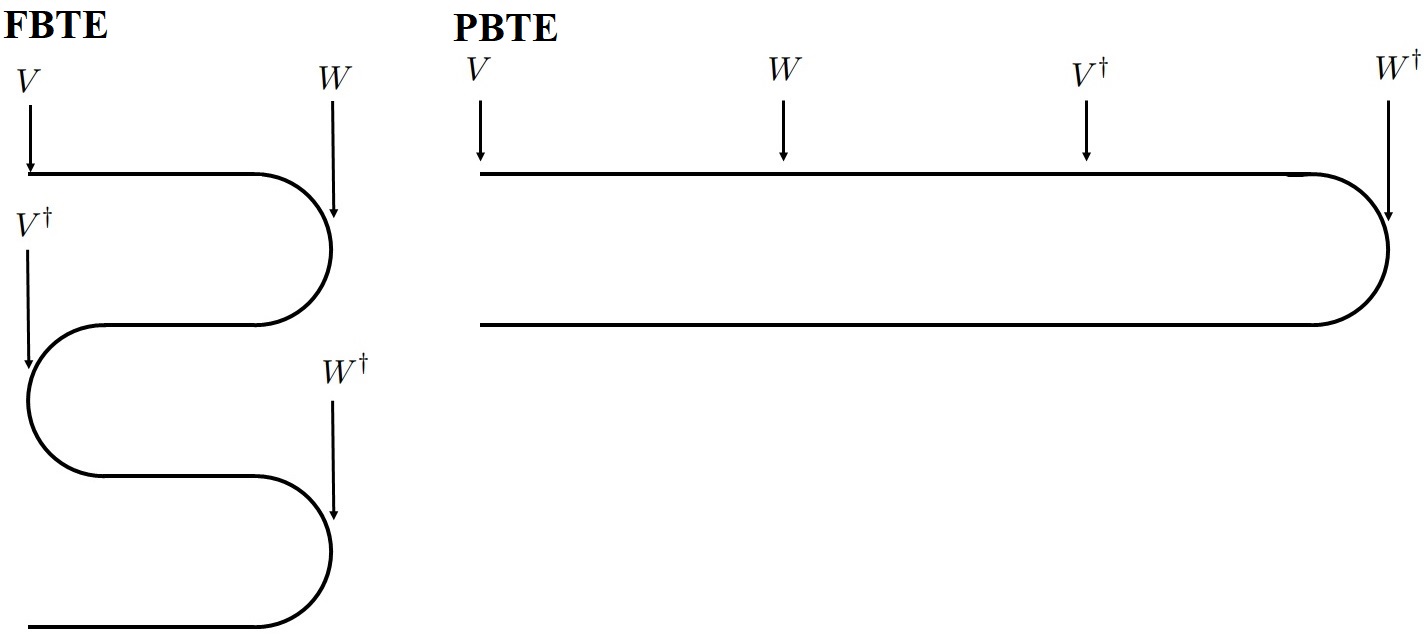}
\caption{Schematic representation of the two considered environment evolutions: the full  (left panel) and the partial (right panel) time reversal case (for details see text). 
In the FBTE case measurements performed on the spin chain, apart from the first one, happen at points, at which the evolution of the environment is reversed. In the PBTE case kets of the environmental density matrix evolve forward in time (bras backward), all measurements take place on the forward-time branch. The time between subsequent measurements is $t$. \label{fig1:scheme}    }
\centering
\end{figure}

In order to derive expressions for OTOCs in both considered cases let us analyze the following steps of a protocol measuring OTOCs, involving backward time evolution: \\
1. Apply $V$ to $\rho_{SE}$ and perform forward time evolution: In the FBTE as well as the PBTE case given by $U_{SE}$
    \ben
        U_{SE} V \rho_{SE} U_{SE}^{\dagger}. \nonumber
    \een
Here, and in the following, $V = V_S \otimes I_E$ denotes an observable acting trivially on the environment (similarly for $W$) and $\rho_{SE}$ is an initial system-environmental state.
2. Apply $W$ and perform backward time evolution:
        \ben
       &&\text{FBTE:} \; U_{SE}^{\dagger}WU_{SE} V \rho_{SE} U_{SE}^{\dagger}U_{SE} =  U_{SE}^{\dagger}WU_{SE} V \rho_{SE}, \nonumber  \\
    &&\text{PBTE:}\;        U_{S^{\dagger}E}WU_{SE} V \rho_{SE} U_{SE}^{\dagger}U_{SE^{\dagger}}. \nonumber
    \een
At this point the characteristic distinction between FBTE and PBTE case can be seen: in the former unitary operators on the right side of $\rho_{SE}$ form identity operator, in contrary to the latter. Repetition of the above steps for $V^\dagger$ and $W^\dagger$ leads to 
          \ben
    \label{eq:OTOC_full_TR_derivation}
  \text{FBTE:} \;  &&U_{SE}^{\dagger}  W^{\dagger} U_{SE} V^{\dagger} U_{SE}^{\dagger}WU_{SE} V \rho_{SE} U_{SE}^{\dagger}U_{SE} = \nonumber\\&&   U_{SE}^{\dagger}  W^{\dagger} U_{SE} V^{\dagger} U_{SE}^{\dagger}WU_{SE} V \rho_{SE}, 
  \\
    \label{eq:OTOC_partial_TR_derivation}
      \text{PBTE:}\; &&U_{S^{\dagger}E} W^{\dagger}U_{SE}V^{\dagger}U_{S^{\dagger}E}WU_{SE} V \rho_{SE} U_{SE}^{\dagger}U_{SE^{\dagger}}  U_{SE}^{\dagger} U_{SE^{\dagger}}. \nonumber  \\
    \een
By taking trace of Eq. (\ref{eq:OTOC_full_TR_derivation}) and Eq. (\ref{eq:OTOC_partial_TR_derivation}) we obtain 
\ben
\label{eq:OTOCFBTE}
&&F^{OS}_t(V,W)  = \\ &&Tr\left( U_{SE}^{\dagger} W^{\dagger} U_{SE} V^{\dagger} U_{SE}^{\dagger} W U_{SE} V \rho_{SE} \right), \nonumber
\een
for the FBTE case, and
\ben
\label{eq:OTOCPBTE}
&&F^{OS}_t(V,W)  = \\ &&Tr \left( U_{SE}^{\dagger}U_{SE^{\dagger}} U_{SE}^{\dagger}  W^{\dagger} U_{SE} V^{\dagger} U_{S^{\dagger}E} W U_{SE} V \rho_{SE}  \right), \nonumber
\een
for the PBTE case. Let us mention that PBTE scheme has been considered previously only for pure states \cite{YHS2018-Resiliance} so Eq. (\ref{eq:OTOCPBTE}) is the first result of this paper. Note also that the presented reasoning can be straightforwardly applied to higher order OTOCs, i.e. the ones involving more stages, at which evolution of the system or the environment is reversed and more measurements in between are made.

\section{A spin chain case study}
\label{sec:OTOC_spinchain}
In what follows we will focus on a model of a spin-$1/2$ chain, whose sites couple linearly to environments consisting of harmonic oscillators. We will assume that the environments are independent so that there is no coupling between them.   Therefore, the considered class of Hamiltonians is of a form: 
$H_{SE} = H_S + H_E + H_{S:E},$ where $H_S$ is a general $N$-site spin-$1/2$ chain Hamiltonian, $H_E$ describes the environment, which consist of independent collections of harmonic oscillators (one group for each site of the chain) 
\ben
H_E = \sum_{k=0}^{N-1} \sum_{j=0}^{M_N-1}  \omega_{k,j} a^{\dagger}_{k,j} a_{k,j},
\een
and $H_{S:E}$ describes linear coupling between each site and its environment 
\ben
H_{S:E} = \sum_{k=0}^{N-1} \lambda_k \sigma_{z,k} \otimes \sum_{j=0}^{M_N-1} C_{k,j} \left( a^{\dagger}_{k,j} + a_{k,j} \right).
\een
With the help of spin coherent states \cite{RadcliffeSpinCoherent, KlauderPathIntegrals,Kochetov, Kirchner2010}, one can apply the usual path-integral reasoning and formulate expression for OTOCs as a sum over all possible trajectories in the spin phase space.  The details can be found in Appendix \ref{sec:appecolution}, here only main steps of the derivation are presented. We start by using resolution of identity in the basis of spin-coherent states, which in the case of a $N$-site chain reads
\begin{equation}
I =\prod_{k=0}^{N-1}  \int \frac{d \pmb{z}_k d \pmb{z}_k^*}{\pi(1+|\pmb{z}_k|^2)^2}  \ket{\pmb{z}_k} \bra{\pmb{z}_k} \equiv \int d(\pmb{z},\pmb{z}^*)  \ket{\pmb{z}} \bra{\pmb{z}},
\end{equation}
where $\ket{\pmb{z}_k}$ is spin coherent state of the $k$-th site of the chain $\ket{\pmb{z}_k} \equiv \frac{1}{\sqrt{1+|z_k|^2}} e^{\sigma^+_k z_k}\ket{0_k}$ and $\sigma^+_k=\sigma_{x,k}+i\sigma_{y,k}$ \cite{RadcliffeSpinCoherent}. This resolution of identity is inserted before and after each unitary operator in Eqs. (\ref{eq:OTOCFBTE}) and (\ref{eq:OTOCPBTE}). As a result, in those expressions one needs to deal with terms of the form $\bra{\pmb{z_{F_i}}}U_{SE}\ket{\pmb{z_{I_i}}}$, which act on the Hilbert space of the environment. In Appendix \ref{sec:appecolution} we show that they   may be represented as
\begin{eqnarray}
 &&\bra{\pmb{z_{F_i}}}U_{SE}\ket{\pmb{z_{I_i}}}=\int_{\pmb{z_{I_i}}}^{ \pmb{z_{F_i}}} d(\pmb{z}, \pmb{z}^*) e^{\Gamma\left[ \pmb{z}, \pmb{z}^*\right] + iS \left[ \pmb{z}, \pmb{z}^* \right]} \times \nonumber \\&& e^{-i\sum_kH_{E,k} t} e^{-i\sum_k\xi_{k;t}t[\pmb{z}] } D\left( \sum_k\chi_{k,t} [\pmb{z}] \right),
\end{eqnarray}
where the precise form of the action  and surface terms is provided by Eqs. (\ref{eq:action}) and (\ref{eq:boundary}) respectively,  $D(\sum_k\chi_{k}) \equiv e^{\sum_k\chi_{k} a_k^\dagger - \chi_{k}^* a_k }$ is the displacement operator, which argument is given by Eq. (\ref{eq:disparg}), and the phase factor $\xi_{k;t}t[\pmb{z}]$ is given by Eq. (\ref{eq:phasearg}). Subsequently we assume product initial system-environment state i.e. $\rho_{SE} = \rho_S \otimes \bigotimes_{k=0}^{N-1} \rho_{E,k}$, where $\rho_{E,k} = \frac{e^{- \beta  H_{E,k}}}{Tr(e^{- \beta H_{E,k})}}$. The environmental degrees of freedom can be traced out analytically, for details of derivation we refer the interested reader to Appendix \ref{sec:appecolution}. One arrives at the expression, in which the interaction of the environment on the system is captured by the Feynman-Vernon influence functional
\ben
F_t^{OS}(V,W) =&&
\int
d \left( \pmb{Z},\pmb{Z}^* \right)e^{\Gamma\left[\pmb{Z},\pmb{Z}^*  \right]+iS\left[\pmb{Z},\pmb{Z}^* \right]}
\times \nonumber \\ && F_t\left[\mathcal{\pmb{Z}},\mathcal{\pmb{Z}}^* \right]  e^{- \Phi_t \left[\mathcal{\pmb{Z}},\mathcal{\pmb{Z}}^*  \right]},
\een
where, in order to keep formulas concise,   $\mathcal{\pmb{Z}}$ is an abbreviation for all variables of the problem i.e. $ \mathcal{\pmb{Z}} \equiv \pmb{z_1}, \pmb{z_2}, \pmb{z_3}, \pmb{z_3} $, bold stands for a vector   e.g. $\pmb{z_1} \equiv(\pmb{z_1}_{,1}. \dots, \pmb{z_1}_{,N})$, formulas for action $S\left[\pmb{Z},\pmb{Z}^* \right]$  and surface $\Gamma\left[\pmb{Z},\pmb{Z}^*  \right]$  terms are presented in Appendix as they are not important in further considerations,  $F_t\left[ \mathcal{\pmb{Z}},\mathcal{\pmb{Z}}^*\right]$  denotes elements of a closed system OTOC expressed in the basis of coherent states 

\ben
F_t\left[ \mathcal{\pmb{Z}},\mathcal{\pmb{Z}}^* \right]=&&W^*(\pmb{z_{I_4}}^*, \pmb{z_{F_3}})V^*(\pmb{z_{I_3}}^*, \pmb{z_{F_2}}) \nonumber \times \\ &&W(\pmb{z_{I_2}}^*, \pmb{z_{F_1}})   (V\rho_S)(\pmb{z_{I_1}}^*, \pmb{z_{F_4}}), 
\een
with $W(\pmb{z^*}, \pmb{z'}) \equiv \bra{\pmb{z}} W\ket{\pmb{z'}}$ and $ \Phi_t \left[\mathcal{\pmb{Z}},\mathcal{\pmb{Z}}^* \right]$ is the influence phase, which explicit form will be presented shortly. Due to the form of the system-environment Hamiltonian, the influence phase consists of contributions coming from individual baths
\ben
    \Phi_t \left[\mathcal{\pmb{Z}},\mathcal{\pmb{Z}}^* \right] = &&\sum_k \Phi_{k;t} \left[\mathcal{\pmb{Z}},\mathcal{\pmb{Z}}^* \right].
\een
It will prove useful to express the results in terms of the usual influence functional obtained for a bosonic thermal bath \cite{KleinertBook}:
\ben
    \Phi^{B}_t[z,z'] = &&\int_{0}^{t} dt' \int_0^{t'} dt'' \left( z(t') - z'(t') \right) \times \nonumber \\  &&\left( \xi_J(t'-t'') z(t'')  - \xi_J^*(t'-t'') z'(t'') \right). 
\een
For convenience, we passed to the description of the environment in terms of the bath correlation function
\ben
    \xi_J(t) = \int_0^{\infty} d \omega && J(\omega) \times  \\ && \left( \coth \left(\frac{\beta \omega}{2} \right) \cos \left( \omega t  \right)+ i \sin\left( \omega t  \right)\right), \nonumber
\een
expressed with the help of a spectral density $J(\omega) = \sum_j C^2_j \delta(\omega-\omega_j)$ encapsulating details of the coupling between the system and the environment.  

In the FBTE case, when the system and the environment jointly undergo forward and backward time evolution, calculation (see Appendix \ref{sec:appecolution} for details) results in the following influence phase

\ben
\label{eq:IF_phase_FBTE}
    \Phi_{k;t} && \left[ \mathcal{\pmb{Z}},\mathcal{\pmb{Z}}^* \right] = \\ &&\Phi^B_t \left[ n_z[\pmb{z_{1}}_{,k}], n_z[\pmb{z_{2}}_{,k}] \right] + \Phi^B_t \left[ n_z[\pmb{z_{3}}_{,k}], n_z[\pmb{z_{4}}_{,k}] \right]+  \nonumber \\
      &&\int_0^{t} dt' \int_0^{t'} dt'' \left(  n_z[\pmb{z_{1}}_{,k}(t')] - n_z[\pmb{z_{2}}_{,k}(t')] \right)  \nonumber \times \\  &&\xi_{J_k}(t'-t'') \left(n_z[\pmb{z_{3}}_{,k}(t'')] - n_z[\pmb{z_{4}}_{,k}(t'')] \right)+  \nonumber \\
      &&\int_0^{t} dt' \int_0^{t'} dt'' \left(  n_z[\pmb{z_{3}}_{,k}(t')] - n_z[\pmb{z_{4}}_{,k}(t')] \right)  \nonumber \times \\  &&\xi_{J_k}(t'-t'') \left(n_z[\pmb{z_{1}}_{,k}(t'')] - n_z[\pmb{z_{2}}_{,k}(t'')] \right), \nonumber
\een
where $n_z[\pmb{z}_{k}]=\frac{1-|\pmb{z}_{k}|^2}{1+|\pmb{z}_{k}|^2}$.  The above expression can be understood in the following way. Two first terms of the influence phase are essentially identical to the standard influence phase for the spin-boson problem. They stem from two pairs of forward-backward time branches (the first and the last one, respectively) in the left panel of Figure \ref{fig1:scheme}. However, those branches are not independent, what is manifested by the last two terms of the influence phase.

In the PBTE case we find (see Appendix \ref{sec:appecolution} for details)  that the influence phase is of a form 
\ben
\label{eq:IF_phase_PBTE}
    \Phi_{k;t}  \left[\mathcal{\pmb{Z}},\mathcal{\pmb{Z}}^* \right] =&& \\ \nonumber 
    && \Phi^B_{k;3t} \left[  n_z\left[ \sum_{r=1}^3 \Pi_{(r-1)t,rt} \pmb{z_{r}}_{,k}\right], n_z[\pmb{z_{4}}_{,k}] \right],
\een 
where is $\Pi_{t_i,t_f}$ a window function defined as a difference of Heaviside step functions $\Pi_{t_i,t_f}f(t')= (\theta(t'-t_i)-\theta(t'-t_f))f(t')$. Comparing the above expression with the standard influence phase for the spin-boson problem we see that on the forward-time branch the external force is composed of three independent terms, which correspond to a paths taken by the spin-chain between measurements (cf. right panel of the Figure \ref{fig1:scheme}).

The influence functionals for the FBTE and PBTE cases have the same proprieties as in the standard Feynman-Vernon formalism \cite{FV1963, GJKK+1996-book}. Most importantly, for both considered cases, the absolute value of the influence functional is bounded from above by one $\left|e^{- \Phi_t \left[\mathcal{\pmb{Z}},\mathcal{\pmb{Z}}^*  \right]}\right| \leq 1$, what is most easily seen from Eqs. (\ref{eq:if_full_TR_derivation}), (\ref{eq:OTOC_partial_TR_IF}) in Appendix. This implies that the absolute value of the open system OTOCs will be in general smaller than that of the corresponding closed quantum systems. One can understand this fact in the following way: In an open system spread of the quantum information, as measured by OTOCs, is faster than that in a corresponding closed system. This is because, in the former, there are additional degrees of freedom provided by the environment, which become correlated with the system degrees of freedom via system-environment interactions.

Eqs. (\ref{eq:IF_phase_FBTE}) and (\ref{eq:IF_phase_PBTE}) are the main contribution of this paper. They allow to compare decoherence scenarios corresponding to FBTE and PBTE schemes, and describe these processes in terms of the microscopic parameters of a considered model.

\section{Applications}
\label{sec:applications}
In this Section, for the sake of presentation clarity, we will assume a uniform coupling strength between sites of the chain and their respective environments as well as the same spectral density for all environments. Our aim is to obtain a bound on open system OTOCs. Let us assume that we are in a regime, in which dephasing dominates over dissipation. In such a case the imaginary part of the influence functional can be neglected \cite{GJKK+1996-book}, what results in purely dephasing channel. Rate of dephasing can be related to the microscopic description of the considered model. First of all, noting that $|n_z[\pmb{z}_{k}(t')]|\leq \frac{1}{2}$, one sees that, in the most destructive case, difference between respective spin trajectories in Eqs. (\ref{eq:IF_phase_FBTE}, \ref{eq:IF_phase_PBTE}) is equal to $1$. As a result, for the FBTE the following bound follows
\ben
 &&|F_t^{OS}| \geq |F_t| e^{- 4\lambda^2 N\int_{0}^{t} dt' \int_0^{t'} dt'' Re \xi_J(t'-t'')},
 \label{eq:FBTELowerBound}
\een
whereas for the PBTE case it reads
\begin{equation}
	 |F_t^{OS}| \geq |F_t| e^{- \lambda^2 N\int_{0}^{3t} dt' \int_0^{t'} dt'' Re \xi_J(t'-t'')}.
	 \label{eq:PBTELowerBound}
\end{equation} 
As an illustration, let us consider a spectral density of a form  $J(\omega) = \frac{\omega^s}{\Lambda^{s-1}}e^{-\omega/\Lambda}$,  known from the spin-boson problem \cite{LCDF+1987-DynamicsDissipativeTwoLevel}. For $s=1$ the relevant integrals can be evaluated, if we assume the low temperature limit, which is determined by the cut-off: $k_B T \ll \Lambda$. In the FBTE case we find
\ben
\label{eq:s1}
\int_0^t dt' \int_0^{t'} dt''Re \xi_J(t'-t'')  = \ln \left[ \sqrt{1+\Lambda^2 t^2} \frac{\sinh (t/\tau_T)}{t/\tau_T} \right], \nonumber \\
\een
where $\tau_T=\frac{1}{\pi k_BT}$ is thermal time. In the case $s>1$, the integrals can be computed analytically for all temperature regimes:
\ben
\label{eq:sg2}
&&\frac{\beta^{s-1} \Lambda^{s-1}}{\Gamma(s-1)}\int_0^t dt' \int_0^{t'} dt''Re \xi_J(t'-t'') \nonumber = \\ &&\zeta\left(s-1,\frac{1}{\Lambda \beta}\right)+   \zeta\left(s-1,1+\frac{1}{\Lambda \beta}\right)- \nonumber \\&& \frac{1}{2}\bigg[ \zeta\left(s-1,\frac{1}{\Lambda \beta}+i\frac{t}{\beta}\right) +  \zeta\left(s-1,1+\frac{1}{\Lambda \beta}+i\frac{t}{\beta}\right) \nonumber  \\ &&+c.c \bigg] , \; \text{for} \; s \neq 2
\een
where $\zeta(p,a)$ is Hurwitz zeta function\cite{NIST}
\ben
\zeta(p,q) = \sum_{n=0}^{\infty} \frac{1}{(q+n)^p},
\een
$\Gamma(s)$ stands for Euler Gamma function, $\beta \equiv \frac{1}{k_BT}$, and $c.c.$ denotes complex conjugation. The case $s=2$ requires a separate treatment, the resulting expression is similar to the above one, with Hurwitz zeta functions replaced by Digamma functions \cite{NIST}
$
\psi(q) = \frac{d}{d q} \ln \Gamma(q)$. 
The analytical expressions in the PBTE case are obtained after a substitution  $t \rightarrow 3t$ in Eqs. (\ref{eq:s1}, \ref{eq:sg2}). 

\begin{figure}[t!]
	
	\includegraphics[scale=0.27]{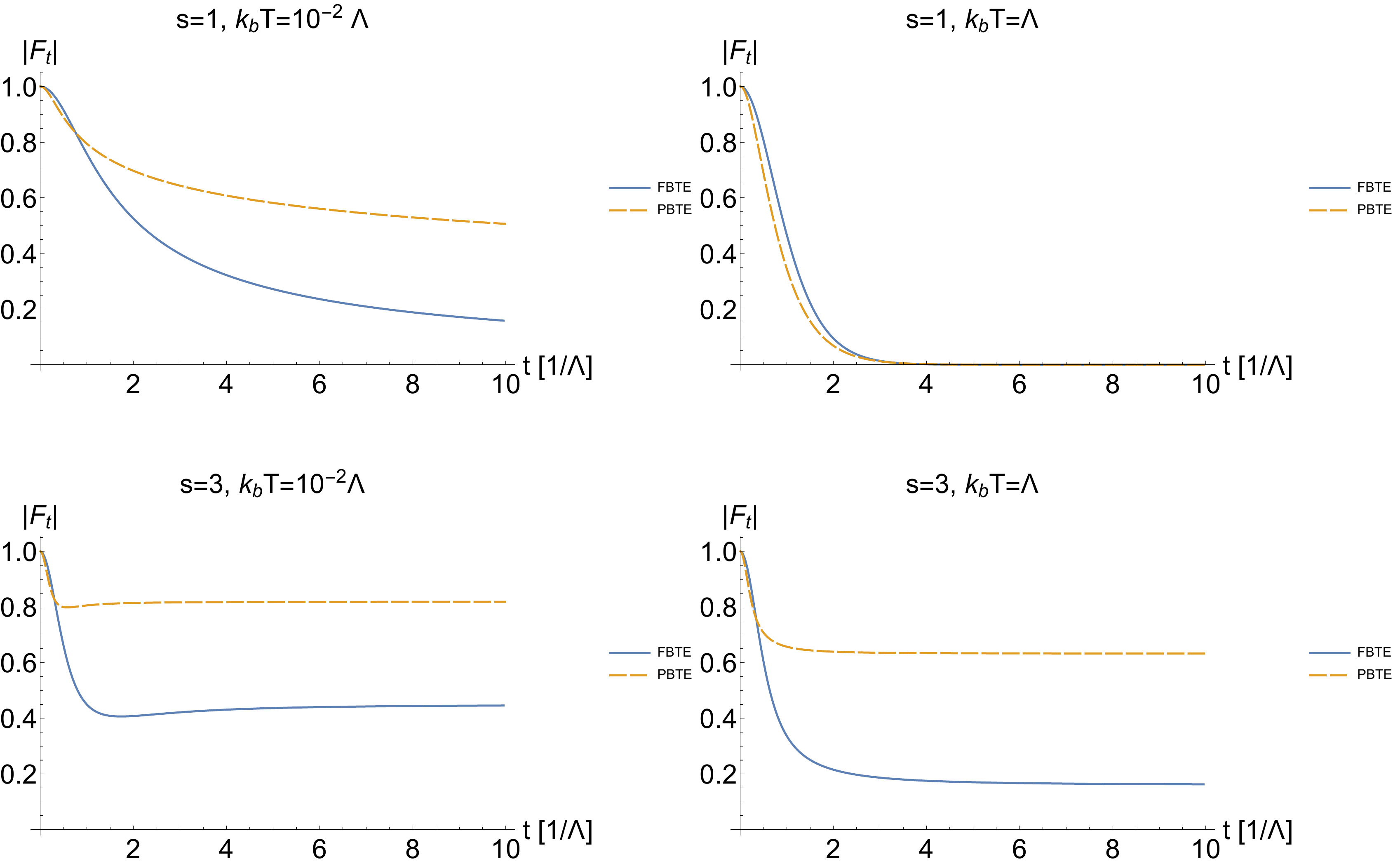}
	\caption{Lower bound on open system OTOC in the pure dephasing regime for FBTE case (c.f. Eq. (\ref{eq:FBTELowerBound})) - solid line, and PBTE case  (c.f. Eq. (\ref{eq:PBTELowerBound})) - dashed line. The left panels show results in the case of low temperature regime ($k_BT=10^{-2} \Lambda$): $s=1$ -- upper one and $s=3$ -- lower one. The right panels show results in the case of intermediate temperature regime $k_BT= \Lambda$: $s=1$ -- upper one and $s=3$ -- lower one. The plots were done for a chain consisting of $N$=20 sites and $\lambda$ = 0.1.  \label{fig2:graph}    }
	\centering
\end{figure}

The performance of the bound for FBTE and PBTE case is illustrated in Figure \ref{fig2:graph}. In the pure dephasing scenario, PBTE case leads to a tighter bound on open system OTOCs, especially in the case of the superohmic spectral density (i.e. for $s>1$). The possible explanation of this fact may be related to the phenomenon of non-Markovianity (for general introduction see e.g.\cite{BreuerRevMod}). For the spin-boson model it is known that super-ohmic spectral densities lead to non-Markovian evolution \cite{BreuerPRA, BylickaPRA}: A depahsing qubit can regain coherences previously lost to the environment. OTOCs aim at measuring spread of quantum correlations across system degrees of freedom, which in open systems is enhanced by the presence of the environment. This unwanted enhancement can be suppressed if some quantum information lost to the environment will flow back to the system, due to non-Markovian memory effects. It is plausible that such a back flow may be more significant in PBTE compared to the FBTE case, in which evolution of the environment is also reversed.  It would be interesting to further investigate this issue.

It would be desirable to generalize the above bound in order to account for dissipation. In the most straightforward way this can be done by application of a similar reasoning to that presented in \cite{MSHP2017-SpinBosonError}. Estimation of the subsequent terms in Taylor expansion of the influence functional leads to a bound on $|\Delta F_t(V,W)| \equiv |F_t(V,W)-F^{OS}_t(V,W)|$, i.e.  the difference between an OTOC and its open system counterpart. It reads 
\begin{equation}
\frac{|\Delta F_t(V,W)|}{|F_t(V,W)|} \leq  e^{- 4\lambda^2 N\int_{0}^{t} dt' \int_0^{t'} dt'' | \xi_J(t'-t'')|} -1 
\end{equation}
 for FBTE case (a similar expression can be found for PBTE case). However, in general we have $|\Delta F_t(V,W)|/|F_t(V,W)| \leq 1 $, and numerical simulations show that the right hand side of the above inequality quickly exceeds 1, what makes the above bound not a useful one. To improve tightness of the bound a more careful treatment is required, e.g. one using a suitably modified version of non-interacting blip approximation \cite{LCDF+1987-DynamicsDissipativeTwoLevel}.
\section{Summary and outlook} \label{sec:discussion}
In this work we applied Feynman-Vernon influence functional technique to study open system OTOCs. We considered two possible backward time evolution schemes -- in the first one the evolution of the environment was reversed whereas in the second is was not. We derived expressions for open systems OTOCs in both cases. Subsequently we considered the model of a one-dimensional spin-$1/2$ chain interacting with bosonic environment and computed the influence phase for both scenarios. The influence phase was used to derive bounds on open system OTOCs. The behavior of the bounds was analyzed for the spectral density known from the spin-boson model.  

It would be interesting to extend the present study for higher spins. This requires careful treatment, as it has been shown that, for spins $s>\frac{1}{2}$, the spin-coherent path integrals are not well defined \cite{WG2011-BreakdownCoherentPathIntegral}. However, the resolution of this problem has been also proposed \cite{KKK-CoherentPathIntegralContinuum}, what may open a path for higher-spin extension. Moreover, a potential future research direction concerns deriving a master equation for open system OTOCs using the calculated influence phase. Although at present no closed expression for a master equation corresponding to the spin-boson problem is known \cite{F2017-SpinBosenME}, the results obtained here will have a very similar structure to those for a bosonic central system system. In such a case the standard techniques of deriving master equation from influence phase should apply. This problem will be studied elsewhere.        
\begin{acknowledgments}
Helpful remarks from E. Aurell, J. Cotler, and F. Wilczek  are acknowledged. This work was supported by the
the European Research Council under grant 742104.
\end{acknowledgments}
\bibliography{bibliography}

\newpage
\appendix

\section{Evolution of the environment}
\label{sec:appecolution}
Here we derive expression for the evolution operator of the environmental degrees of freedom. We start by expressing the full evolution operator in the basis of spin-coherent states:
\begin{equation}
    \bra{\pmb{z_{F_i}}}U_{SE}\ket{\pmb{z_{I_i}}}=  \bra{\pmb{z_{F_i}}}\lim_{N \rightarrow \infty }\prod_{n=1}^{N}e^{-i(H_S+H_E+H_{S:E})\Delta t}\ket{\pmb{z_{I_i}}},
\end{equation}
where $\Delta t \equiv \frac{t}{N}$. In the next step, we insert resolution of identity, expressed in terms of spin coherent states
\begin{equation}
    I =\prod_k  \int \frac{d \pmb{z}_k d \pmb{z}_k^*}{\pi(1+|\pmb{z}_k|^2)^2}  \ket{\pmb{z}_k} \bra{\pmb{z}_k} \equiv \int d(\pmb{z},\pmb{z}^*)  \ket{\pmb{z}} \bra{\pmb{z}},
\end{equation}
between subsequent terms of the product

\ben
\label{eq:pathintegral1}
  &&\bra{\pmb{z_{F_i}}}U_{SE}\ket{\pmb{z_{I_i}}} = \nonumber \\&& \bra{\pmb{z_{F_i}}}\lim_{N \rightarrow \infty }\prod_{n=1}^{N}  \left( \int \int d(\pmb{z_{n+1}}, \pmb{z_{n+1}}^*)d(\pmb{z_{n}}, \pmb{z_{n}}^*) \right. \nonumber\\&&  \left. \ket{\pmb{z_{n+1}}}  
  \bra{\pmb{z_{n+1}}} e^{-i(H_S+H_E+H_{S:E})\Delta t} \ket{\pmb{z_n}} \bra{\pmb{z_n}}  \right)\ket{\pmb{z_{I_i}}} 
\een
We focus on a single term
\begin{eqnarray}
\label{eq:pathintegral2}
&&\bra{\pmb{z_{n+1}}} e^{-i(H_S+H_E+H_{S:E})\Delta t} \ket{\pmb{z_n}} \approx \nonumber \\&&  \bra{\pmb{z_{n+1}}} \left(1-i(H_S+H_E+H_{S:E})\Delta t \right) \ket{\pmb{z_n}} = \nonumber \\ \nonumber  &&\< \pmb{z_{n+1}}\left. \right| \pmb{z_n} \> \times \nonumber \\ &&\left(1-i(H_S(\pmb{z_{n+1}},\pmb{z_{n}})+H_E+H_{S:E}(\pmb{z_{n+1}},\pmb{z_{n}}))\Delta t \right)  \approx \nonumber \\&& \< \pmb{z_{n+1}}\left. \right| \pmb{z_n} \> e^{-iH_S(\pmb{z_{n+1}},\pmb{z_{n}}))\Delta t}e^{-i(H_E+H_{S:E}(\pmb{z_{n+1}},\pmb{z_{n}}))\Delta t} \nonumber \\
\end{eqnarray}
Our final aim is to take the limit $N \rightarrow \infty$, when the states will become close to each other, i.e. $\Delta \pmb{z_{n}}_{,k} \equiv \pmb{z_{n+1}}_{,k} - \pmb{z_{n}}_{,k} = O(\Delta t)$. In such a case the scalar product of spin coherent states becomes \cite{Kochetov}
\ben
\label{eq:spinscalarapprox}
&&\< \pmb{z_{n+1}}\left. \right| \pmb{z_n} \> = \prod_k\frac{1+\pmb{z_{n+1}}_{,k}\pmb{z_n}_{,k}^*}{\sqrt{1+|\pmb{z_{n+1}}_{,k}|^2}\sqrt{1+|\pmb{z_{n}}_{,k}|^2}}\nonumber \\ \nonumber && \approx \prod_k \exp \left[ \frac{\pmb{z_{n}}_{,k} \frac{\Delta \pmb{z_{n}}_{,k}^*}{\Delta t} - \pmb{z_{n}}_{,k}^* \frac{\Delta \pmb{z_{n}}_{,k}}{\Delta t}   }{1+|\pmb{z_n}_{,k}|^2} \Delta t \right]  \equiv \\&& \exp \left[ \frac{\pmb{z_{n}} \frac{\Delta \pmb{z_{n}}^*}{\Delta t} - \pmb{z_{n}}^* \frac{\Delta \pmb{z_{n}}}{\Delta t}   }{1+|\pmb{z_n}|^2} \Delta t \right],
\een
and the elements of spin operators entering the Hamiltonian are replaced by \cite{Kirchner2010}
\ben
&&\frac{\bra{\pmb{z_{n+1}}_{,k}} \sigma_{x,k} \ket{\pmb{z_{n}}_{,k}}}{\< \pmb{z_{n+1}}_{,k} \left. \right|   \pmb{z_{n}}_{,k} \>} = \frac{Re \pmb{z_{n}}_{,k} }{1+|\pmb{z_{n}}_{,k}|^2}\equiv n_x[\pmb{z_{n}}_{,k}] \nonumber \\
&&\frac{\bra{\pmb{z_{n+1}}_{,k}} \sigma_{y,k} \ket{\pmb{z_{n}}_{,k}}}{\< \pmb{z_{n+1}}_{,k} \left. \right|   \pmb{z_{n}}_{,k} \>} = \frac{Im \pmb{z_{n}}_{,k} }{1+|\pmb{z_{n}}_{,k}|^2}\equiv n_y[\pmb{z_{n}}_{,k}] \nonumber \\
&&\frac{\bra{\pmb{z_{n+1}}_{,k}} \sigma_{z,k} \ket{\pmb{z_{n}}_{,k}}}{\< \pmb{z_{n+1}}_{,k} \left. \right|   \pmb{z_{n}}_{,k} \>} = \frac{|\pmb{z_{n}}_{,k}|^2 -1 }{1+|\pmb{z_{n}}_{,k}|^2}\equiv n_z[\pmb{z_{n}}_{,k}] 
\label{eq:spin_operators_replacement}
\een
Using Eq. (\ref{eq:spinscalarapprox}), Eq. (\ref{eq:pathintegral2}) and Eq (\ref{eq:spin_operators_replacement}), we can formally take the limit of Eq. (\ref{eq:pathintegral1}) :
\ben
\label{eq:controlunitary}
&&\int_{\pmb{z_{I_i}}}^{\pmb{z_{F_i}}} d(\pmb{z_i}, \pmb{z_i}^*) e^{\Gamma\left[ \pmb{z_i}, \pmb{z_i}^*\right] + iS \left[ \pmb{z_i}, \pmb{z_i}^* \right]} \mathcal{T} e^{-i\int_0^t dt'(H_E +  H_{S:E}[\pmb{z_i}(t')])} = \nonumber \\&&\int_{\pmb{z_{I_i}}}^{ \pmb{z_{F_i}}} d(\pmb{z_i}, \pmb{z_i}^*) e^{\Gamma\left[ \pmb{z_i}, \pmb{z_i}^*\right] + iS \left[ \pmb{z_i}, \pmb{z_i}^* \right]} \times \nonumber \\&& e^{-i\sum_kH_{E,k} t} e^{-i\sum_k\xi_{k;t}t[\pmb{z_i}] } D\left( \sum_k\chi_{k,t} [\pmb{z_i}] \right).  
\een
The action $S \left[ \pmb{z_i}, \pmb{z_i}^* \right]$ and the boundary term $\Gamma\left[ \pmb{z_i}, \pmb{z_i}^*\right]$ are given by \cite{Kochetov}:
\begin{equation} 
\label{eq:action}
\begin{aligned} S\left[ \pmb{z_i},  \pmb{z_i}^{*}\right] \equiv\int_{0}^{t} \mathrm{d} t^{\prime}\left(\frac{i}{2} \frac{ \pmb{z_i}^{*}\left(t^{\prime}\right) \dot{ \pmb{z_i}}\left(t^{\prime}\right)-\dot{ \pmb{z_i}}^{*}\left(t^{\prime}\right)  \pmb{z_i}\left(t^{\prime}\right)}{1+\left| \pmb{z_i}\left(t^{\prime}\right)\right|^{2}}\right.\\-H\left( \pmb{z_i}\left(t^{\prime}\right),  \pmb{z_i}^{*}\left(t^{\prime}\right)\right) \bigg) \end{aligned},
\end{equation}
and
\begin{equation}
\label{eq:boundary} 
\Gamma\left[ \pmb{z_i},  \pmb{z_i}^{*}\right] \equiv \frac{1}{2} \log \left(\frac{\left(1+ \pmb{z_i}^{*}(0)  \pmb{z_{I_i}}\right)\left(1+ \pmb{z_{F_i}}^{*}  \pmb{z_i}(t)\right)}{\left(1+\left| \pmb{z_{I_i}}\right|^{2}\right)\left(1+\left| \pmb{z_{F_i}}\right|^{2}\right)}\right),
\end{equation}
respectively.  Moreover, in the above expression $D(\sum_k\chi_{k}) \equiv e^{\sum_k\chi_{k} a_k^\dagger - \chi_{k}^* a_k }$ is multimode displacement operator, which argument reads
 \ben
 \label{eq:disparg}
    \chi_{k;t}[\pmb{z}] = -i\sum_j  \int_0^{t} dt' C_{k,j} e^{i \omega t'} n_z[\pmb{z}_{k}(t')],
\een
where $n_z[\pmb{z}_{k}]=\frac{1-|\pmb{z}_{k}|^2}{1+|\pmb{z}_{k}|^2}$, and the phase is
\ben
\label{eq:phasearg}
     &&\xi_{k;t}[\pmb{z}]= \\ &&\int_0^{\infty} J_k(\omega) \int_0^t dt' \int_0^{t'} dt'' n_z[\pmb{z}_{k}(t')] n_z[\pmb{z}_{k}(t'')] \sin \left(\omega \left(t'-t'' \right) \right) \nonumber.
\een
The above expression were written using spectral density $J(\omega) = \sum_j C^2_j\delta(\omega-\omega_j) $.

The evolution operator (\ref{eq:controlunitary}) may now be used to obtain the path integral representation for the open system OTOCs. In the FBTE case we have that
\begin{widetext}
\begin{eqnarray}
\label{eq:FBTE_derivation}
F^{OS}_t(V,W)  = \int
d \left( \pmb{Z},\pmb{Z}^* \right) 	 &&Tr_E\bigg( \bra{\pmb{z_{F_4}}} U_{SE}^{\dagger}\ket{\pmb{z_{I_4}}} \bra{\pmb{z_{I_4}}} W^{\dagger} \ket{\pmb{z_{F_3}}} \bra{\pmb{z_{F_3}}} U_{SE} \ket{\pmb{z_{I_3}}} \times \\&&\bra{\pmb{z_{I_3}}} V^{\dagger} \ket{\pmb{z_{F_2}}} \bra{\pmb{z_{F_2}}} U_{SE}^{\dagger} \ket{\pmb{z_{I_2}}} \bra{\pmb{z_{I_2}}} W \ket{\pmb{z_{F_1}}} \bra{\pmb{z_{F_1}}} U_{SE} \ket{\pmb{z_{I_1}}} \bra{\pmb{z_{I_1}}} V \rho_{SE} \ket{\pmb{z_{F_4}}}  \bigg) = \nonumber \\ \nonumber
 \int
d \left( \pmb{Z},\pmb{Z}^* \right)&& F_t\left[ \mathcal{\pmb{Z}},\mathcal{\pmb{Z}}^* \right]   Tr_E\bigg( \bra{\pmb{z_{F_4}}} U_{SE}^{\dagger}\ket{\pmb{z_{I_4}}}  \bra{\pmb{z_{F_3}}} U_{SE} \ket{\pmb{z_{I_3}}} \bra{\pmb{z_{F_2}}} U_{SE}^{\dagger} \ket{\pmb{z_{I_2}}} \bra{\pmb{z_{F_1}}} U_{SE} \ket{\pmb{z_{I_1}}} \rho_E  \bigg),= \nonumber \\ \nonumber
\int
d \left( \pmb{Z},\pmb{Z}^* \right)&& e^{\Gamma\left[\pmb{Z},\pmb{Z}^*  \right]+iS\left[\pmb{Z},\pmb{Z}^* \right]} F_t\left[ \mathcal{\pmb{Z}},\mathcal{\pmb{Z}}^* \right] e^{-\Phi[\pmb{Z},\pmb{Z}^*]} ,
\end{eqnarray}
\end{widetext}
where   $\mathcal{\pmb{Z}}$ is an abbreviation for all variables of the problem i.e. $ \mathcal{\pmb{Z}} \equiv \pmb{z_1}, \pmb{z_2}, \pmb{z_3}, \pmb{z_3} $, bold stands for a vector   e.g. $\pmb{z_1} \equiv(\pmb{z_1}_{,1}. \dots, \pmb{z_1}_{,N})$, and 
\ben
F_t\left[ \mathcal{\pmb{Z}},\mathcal{\pmb{Z}}^* \right]=&&W^*( \pmb{z_{F_3}},\pmb{z_{I_4}}^*)V^*( \pmb{z_{F_2}},\pmb{z_{I_3}}^*) \nonumber \times \\ &&W( \pmb{z_{F_1}},\pmb{z_{I_2}}^)   (V\rho_S)( \pmb{z_{F_4}},\pmb{z_{I_1}}^*), 
\een
with $W( \pmb{z'},\pmb{z^*},) \equiv \bra{\pmb{z}} W\ket{\pmb{z'}}$. We proceed by inserting  Eq. \ref{eq:controlunitary} into Eq.(\ref{eq:FBTE_derivation}). We find
\begin{eqnarray}
	F_t^{OS}(V,W) =&&
	\int_{\pmb{Z_{I}}}^{\pmb{Z_{F}}^*}
	d \left( \pmb{Z},\pmb{Z}^* \right)e^{\Gamma\left[\pmb{Z},\pmb{Z}^*  \right]+iS\left[\pmb{Z},\pmb{Z}^* \right]}
	\times \nonumber \\ && F_t\left[\mathcal{\pmb{Z}},\mathcal{\pmb{Z}}^* \right]  e^{- \Phi_t \left[\mathcal{\pmb{Z}},\mathcal{\pmb{Z}}^*  \right]},
\end{eqnarray}
where
\begin{eqnarray}
	&&S\left[\pmb{Z},\pmb{Z}^* \right] = S\left[\pmb{z_1},\pmb{z_1}^* \right] - S\left[\pmb{z_2}^*,\pmb{z_2} \right] +S\left[\pmb{z_3},\pmb{z_3}^* \right] - S\left[\pmb{z_4}^*,\pmb{z_4}, \right] \nonumber \\
	&& \Gamma\left[\pmb{Z},\pmb{Z}^* \right] = \Gamma\left[\pmb{z_1},\pmb{z_1}^* \right] + \Gamma\left[\pmb{z_2},\pmb{z_2}^* \right] +\Gamma\left[\pmb{z_3},\pmb{z_3}^* \right] + \Gamma\left[\pmb{z_4},\pmb{z_4}^* \right], \nonumber \\
\end{eqnarray}
and the influence functional is 
\ben
\label{eq:if_full_TR_derivation}
e^{-\Phi[\pmb{Z},\pmb{Z}^*]} = \nonumber\prod_k  &&e^{i \left( \xi_{k;t}[\pmb{z_1}] - \xi_{k;t}[\pmb{z_2}] + \xi_{k;t}[\pmb{z_4}] -  \xi_{k;t}[\pmb{z_4}] \right) } \times \\&&Tr \left(  D^{\dagger}\left( \chi_{k,t} [\pmb{z_4}] \right)D\left( \chi_{k,t}  [\pmb{z_3}] \right) \times \right. \nonumber \\ && D^{\dagger}\left( \chi_{k,t} [\pmb{z_2}] \right)  \left. D\left( \chi_{k,t} [\pmb{z_1}] \right) \rho_{E,k}  \right).
\een
The initial state of the environment is represented in terms of Glauber-Sudarshan $P$ function $\rho_{E,k} = \int d \gamma_k d \gamma_k^* P(\gamma_k) \ket{\gamma_k} \bra{\gamma_k}$. Then a straightforward calculation gives 
\ben
\label{eq:OTOC_partial_FBTE_IF}
e^{-\Phi[\pmb{Z},\pmb{Z}^*]}=&&\int d \gamma_k d \gamma_k^* P(\gamma_k) e^{-   |\sum_{i=1,3} \Delta \chi_{k,t}[\pmb{z_i},\pmb{z_{i+1}}]|^2/2}\nonumber \\
&&e^{2 i\sum_{k=1,3} Im\Delta \chi_{k,t}[\pmb{z_i},\pmb{z_{i+1}}] \gamma_k^* } \nonumber \times \\&& e^{ i\sum_{k=1,3} \left( \Delta \xi_{k,t}[\pmb{z_i},\pmb{z_{i+1}}]  +    Im \chi_{k,t} [\pmb{z_i}] \chi_{k,t} [\pmb{z_{i+1}}]^*\right)} \nonumber \times \\&& e^{i Im \Delta \chi_{k,t}[\pmb{z_3},\pmb{z_{4}}] \Delta \chi_{k,t}[\pmb{z_1},\pmb{z_{2}}]^*    },
\een
where $\Delta \chi_{k,t}[\pmb{z_i},\pmb{z_{i+1}}] \equiv \chi_{k,t}[\pmb{z_i}] - \chi_{k,t}[\pmb{z_{i+1}}]$, and similarly $\Delta \xi_{k,t}[\pmb{z_i},\pmb{z_{i+1}}] \equiv  \xi_{k,t}[\pmb{z_i}] -  \xi_{k,t}[\pmb{z_{i+1}}] $.  Assuming that that environment is initialized as a thermal state, which corresponding $P$ function is of a form $P(\gamma) = e^{-|\gamma|^2/\bar{n}}$ with $\bar{n}$ being mean photon number $\bar{n} = \frac{1}{1-e^{-\beta \omega}}$. In such a case the integral in Eq. (\ref{eq:OTOC_partial_FBTE_IF}) can be computed analytically and the resulting expression is Eq. (\ref{eq:IF_phase_FBTE}) of the main text.

The FBTE case is more familiar form the point of view of standard open system theory: environmental ket-states evolve forward in time, whereas environmental bra-states evolve backward in time. All the measurements take place on the forward time branch. As a result one finds that there is one displacement operator corresponding to the forward path, for which the driving force changes at the measurement times, as well as one corresponding to the backward path. More precisely, one inserts resolution of identity into Eq. (\ref{eq:OTOCPBTE}) 
\begin{widetext}
\begin{eqnarray}
F^{OS}_t(V,W)  = \int
d \left( \pmb{Z},\pmb{Z}^* \right) 	 &&Tr_E\bigg( \bra{\pmb{z_{F_4}}}U_{SE}^{\dagger}\ket{\pmb{z}'} \bra{\pmb{z'}}U_{SE^{\dagger}} \ket{\pmb{z}} \bra{\pmb{z}} U_{SE}^{\dagger} \ket{\pmb{z_{I_4}}} \bra{\pmb{z_{I_4}}}  W^{\dagger} \ket{\pmb{z_{F_3}}} \\&& \bra{\pmb{z_{F_3}}} U_{SE} \ket{\pmb{z_{I_3}}} \bra{\pmb{z_{I_3}}} V^{\dagger} \ket{\pmb{z_{F_2}}} \bra{\pmb{z_{F_2}}} U_{S^{\dagger}E} \ket{\pmb{z_{I_2}}} \bra{\pmb{z_{I_2}}} W \ket{\pmb{z_{F_1}}} \bra{\pmb{z_{F_1}}} U_{SE} \ket{\pmb{z_{I_1}}} \bra{\pmb{z_{I_1}}} V \rho_{SE}\ket{\pmb{z_{F_4}}}  \bigg) = \nonumber \\  \int
d \left( \pmb{Z},\pmb{Z}^* \right)&&F_t\left[ \mathcal{\pmb{Z}},\mathcal{\pmb{Z}}^* \right]   Tr_E\bigg( \bra{\pmb{z_{F_4}}}U_{SE}^{\dagger}\ket{\pmb{z}'} \bra{\pmb{z}'}U_{SE^{\dagger}} \ket{\pmb{z}} \bra{\pmb{z}} U_{SE}^{\dagger} \ket{\pmb{z_{I_4}}} \times \nonumber \\&& \bra{\pmb{z_{F_3}}} U_{SE} \ket{\pmb{z_{I_3}}} \bra{\pmb{z_{F_2}}} U_{S^{\dagger}E} \ket{\pmb{z_{I_2}}} \bra{\pmb{z_{F_1}}} U_{SE} \ket{\pmb{z_{I_1}}} \rho_E \bigg), \nonumber 
 \\ \nonumber \int
d \left( \pmb{Z},\pmb{Z}^* \right)&& e^{\Gamma\left[\pmb{Z},\pmb{Z}^*  \right]+iS\left[\pmb{Z},\pmb{Z}^* \right]} F_t\left[ \mathcal{\pmb{Z}},\mathcal{\pmb{Z}}^* \right]   e^{-\Phi[\pmb{Z},\pmb{Z}^*]}, \nonumber
\end{eqnarray}
where
\begin{eqnarray}
	&&S\left[\pmb{Z},\pmb{Z}^* \right] = S\left[\pmb{z_1},\pmb{z_1}^* \right] + S\left[\pmb{z_2},\pmb{z_2}^* \right] +S\left[\pmb{z_3},\pmb{z_3}^* \right] + \bar{S}\left[\pmb{z_4},\pmb{z_4}^* \right], 
\end{eqnarray}
and
\begin{eqnarray}
\bar{S}\left[\pmb{z_4},\pmb{z_4}^* \right] = \int_{0}^{3t} 	&& \mathrm{d} t^{\prime}\left(\frac{i}{2} \frac{ \pmb{z_i}^{*}\left(t^{\prime}\right) \dot{ \pmb{z_i}}\left(t^{\prime}\right)-\dot{ \pmb{z_i}}^{*}\left(t^{\prime}\right)  \pmb{z_i}\left(t^{\prime}\right)}{1+\left| \pmb{z_i}\left(t^{\prime}\right)\right|^{2}}\right. \\ &&+ \left(\Pi_{0,t} + \Pi_{2t,3t} \right)H\left( \pmb{z_i}\left(t^{\prime}\right),  \pmb{z_i}^{*}\left(t^{\prime}\right)\right) - \Pi_{t,2t} H\left( \pmb{z_i}\left(t^{\prime}\right),  \pmb{z_i}^{*}\left(t^{\prime}\right)\right)  \bigg) \nonumber \\
 \Gamma\left[\pmb{Z},\pmb{Z}^* \right] = &&\Gamma\left[\pmb{z_1},\pmb{z_1}^* \right] +	 \Gamma\left[\pmb{z_2},\pmb{z_2}^* \right] +\Gamma\left[\pmb{z_3},\pmb{z_3}^* \right] + \Gamma^*\left[\pmb{z_4},\pmb{z_4}^* \right]. \nonumber 
\end{eqnarray}
\end{widetext}
Expression for the influence functional is 
\ben
\label{eq:OTOC_partial_TR_IF}
e^{-\Phi[\pmb{Z},\pmb{Z}^*]} = \nonumber\prod_k  &&e^{i \left( \xi_{k;3t}[\pmb{z_1} + \pmb{z_2}+ \pmb{z_3}] -  \xi_{k;t}[\pmb{z_4}] \right) } \times \\&&Tr \left( \rho_{E,k}  D^{\dagger}\left( \chi_{k,t} [\pmb{z_4}] \right) \times  \right. \nonumber \\ &&  \left. D\left( \chi_{k,3t} [\pmb{z_1} + \pmb{z_2} + \pmb{z_3}]  \right)   \right),
\een
where the argument of the forward displacement operator reads
\ben
&&\chi_{k;3t}[\pmb{z_1} + \pmb{z_2}+ \pmb{z_3}] = -i\sum_j\int_0^{3t} dt' C_{k,j}(\omega) \times \nonumber \\&& e^{i \omega t'}  n_z\left[\sum_{r=1}^3\Pi_{(r-1)t,rt}\pmb{z}_{k}(t')\right],
\een
and the corresponding phase is
\ben
&&\xi_{k;3t}[\pmb{z_1} + \pmb{z_2}+ \pmb{z_3}]= \nonumber \\&&  \nonumber \int_0^{3t} dt' \int_0^{t'} dt'' \int_0^{\infty} J_k(\omega)     n_z\left[\sum_{r=1}^3 \Pi_{(r-1)t,rt} \pmb{z_r}_{,k}(t')\right] \times  \\&&  n_z\left[\sum_{r'=1}^3 \Pi_{(r'-1)t,r't}\pmb{z_{r'}}_{,k}(t'')\right] \sin \left(\omega \left(t'-t'' \right) \right).
\een

Evaluation of Eq. (\ref{eq:OTOC_partial_TR_IF}) leads to the following expression
\ben
&&\int d \gamma_k d \gamma_k^* P(\gamma_k) e^{-   |\Delta \chi_{k,t}[\pmb{z_1}+\pmb{z_2}+\pmb{z_3},\pmb{z_{4}}]|^2/2} \nonumber \\&&e^{2 i Im\Delta \chi_{k,t}[\pmb{z_1}+\pmb{z_2}+\pmb{z_3},\pmb{z_{4}}] \gamma_k^* } \times \nonumber \\&& e^{ i \Delta \xi_{k,t}[\pmb{z_1}+\pmb{z_2}+\pmb{z_3},\pmb{z_{4}}]  +    Im \chi_{k,t}[\pmb{z_1}+\pmb{z_2}+\pmb{z_3},\pmb{z_{4}}] \chi_{k,t} [\pmb{z_{4}}]^*}   ,
\een
which, as in the previous case, can be computed for thermal states of the environment leading  to Eq. (\ref{eq:IF_phase_PBTE}) of the main text.

\end{document}